\documentclass[twocolumn]{aastex631}
\usepackage{amsmath}	
\usepackage{amssymb}	
\usepackage{hyperref}
\usepackage{newtxtext,newtxmath}
\usepackage{float}
\usepackage{longtable}
\usepackage{tabularx}
\usepackage{float}
\usepackage{comment}
\usepackage{ragged2e}
\usepackage{booktabs}
\renewcommand{\baselinestretch}{1.10}
\usepackage{todonotes}

\shortauthors{Misra et al.}

\begin{document}

\title{The dependence of Polarization Degree of Blazars on particle energy distributions}

\correspondingauthor{Hritwik Bora}
\email{hritwikbora@gmail.com}

\author[0000-0002-7609-2779]{Ranjeev Misra}
\affiliation{Inter-University Centre for Astronomy and Astrophysics, Post Bag 4, Ganeshkhind, Pune - 411007, India}

\author[0000-0002-1520-057X]{Hritwik Bora}
\affiliation{Department of Physics, Tezpur University, Tezpur-784028, India}

\author[0000-0003-1715-0200]{Rupjyoti Gogoi}
\affiliation{Department of Physics, Tezpur University, Tezpur-784028, India}








\begin{abstract}

We present a formalism to predict the Polarization Degree (PD) for synchrotron emission from particles having a specified energy distribution in the presence of an ordered and random magnetic field configuration.  The broad band spectral energy distribution as well as the X-ray and optical PD data for Mrk 501 have been fitted using the formalism. As reported earlier, we find that for a broken power law particle energy distribution, the PD cannot be explained, unless it is assumed that the higher energy particles experience a higher ordered magnetic field compared to the lower energy ones. On the other hand, a log parabola particle energy distribution can explain the observed higher X-ray PD compared to the optical, even when all the particles experience the same magnetic configuration. We discuss the possibility of distinguishing the two scenarios using future observations.

\end{abstract}


\section{Introduction} \label{sec:intro}

Blazars are Active galactic Nuclei that have jets beamed towards us and their spectral Energy Distribution (SED) can be described by two broad and typically separated components ranging from IR to TeV energies \citep{urry1998multiwavelength,massaro2004log}. The lower energy component is modeled as synchrotron emission of non-thermal electron, while the higher energy one could be due to Inverse Comptonization from the same electron distribution \citep{1992A&A...256L..27D, 1993ApJ...416..458D} or could be due to hadronic process \citep{2013EPJWC..6105003B}. For some blazars, the lower energy synchrotron component dominates the emission in X-rays and these are classified as high-energy peaked blazars or HBLs \citep{ghisellini1997optical, 1998MNRAS.299..433F}.

Recent, Imaging X-ray Polarimetry Explorer (IXPE) observations of HBLs have detected X-ray polarization \citep{2021ApJ...912..129Z, 2022Natur.611..677L, 2023NatAs...7.1245D, 2024ApJ...963L..41H}. The polarization degree in X-rays is found to be significantly higher than those detected in the optical and IR bands. This has been interpreted as the high energy electrons that produce the X-ray emission, are in the presence of a more ordered magnetic field compared to the low energy ones that produce the optical/IR emission. In this scenario, the particle accelerations occur in a shock where the magnetic field is ordered and then the particles diffuse outwards to regions with less ordered magnetic fields. If the diffusion time scale is comparable to the cooling one, then the higher energy particles will exist preferentially near the shock and hence experience a more ordered magnetic field. This is in contrast to the standard assumption made during broad band photon spectral fitting that the emission region is homogeneous and the particles of all energies experience the same magnetic field.

An underlying assumption in the interpretation of polarization is that the particle energy distribution is a power law. Broadband spectral energy distribution (SED) fitting typically assumes a broken power law distribution \citep{Anderhub_2009, Abe_2023, abe2024insightsbroadbandemissiontev} and the assumption is valid if the X-ray and optical/IR producing particles have energy less than the break energy of the power law. This indeed will be the case, especially for soft X-rays (2 - 8 keV) where the polarization has been measured. On the other hand, broadband SED has also been shown to be consistent with intrinsically curved particle distribution such as the log parabola form \citep[e.g.][]{2004A&A...413..489M, 2004A&A...422..103M, Tanihata_2004, tramacere, 2009ApJ...691L..13D, 2015A&A...580A.100S,Sinha_2017}. \citet{10.1093/mnras/stae706} and \citet{tantry2024probingbroadbandspectralenergy} show that for Mrk 501, the broadband SED can be fitted using different particle distributions such as broken power law, log parabola and distributions that arise when the acceleration (or escape) time-scale is energy dependent. The jet powers inferred for these distributions with curvature are significantly lower than the one inferred when the distribution is assumed to be a broken power law.

\cite{2024A&A...685A.117M} report multi-wavelength spectral study of Mrk 501 during the three observations by \textit{IXPE}. For all three observations the X-ray coverage is provided by \textit{Swift}-XRT while \textit{NuSTAR} observations are available for the first two. \cite{2024A&A...685A.117M} report that while the $\gamma$-ray fluxes are similar to those obtained for a typical state of the source, the X-ray fluxes are significantly higher, indicating the low Compton dominance of the source during the \textit{IXPE} observations. In this work, we compute the expected polarization degree as a function of photon energy for synchrotron emission from different particle energy distributions. Thus, this work is limited to when the optical/UV and X-ray emission are due to synchrotron emission from energetic leptons and does not have a significant contribution from any Compton or hadronic processes. We compare the results with one of the simultaneous sets of observations by \textit{NuSTAR} and \textit{Swift} during the \textit{IXPE} observations of Mrk 501. We note that the other set of simultaneous observations have similar spectral and polarization properties while the third one doesn't have \textit{NuSTAR} data. The SED of this set of simultaneous observations have been analyzed by \cite{10.1093/mnras/stae706} to show that they can be represented by different particle distributions. Thus, these observations are ideal to test how particle distributions will effect the polarization degree estimate and how they compare with the observations. 

In the next section, we describe the method to estimate the polarization degree for arbitrary particle distributions in a partially ordered magnetic field. In Section \ref{sec:sec3}, we compare the predictions with the observed polarizations for Mrk 501, while in Section \ref{sec:Discussion} we discuss the results and possible future works.  
\section{Polarization from arbitrary particle distributions}\label{sec:sec2}

The single particle synchrotron photon spectrum from an electron with lorentz factor $\gamma$ in the presence of a  random magnetic field distribution with magnitude $B_R$, can be written as \citep[e.g.][]{2009herb.book.....D},

\begin{equation*}
  N_R (x_R) = \frac{2 \pi e^2}{\sqrt{3} \gamma^2 c h} x_R G(\frac{x_R}{2})
  \label{spec_rand}
\end{equation*}

Here, c is the speed of light, h is the Planck constant, and m$_e$ is the mass of the electron.

where,
\begin{equation*}
    G(x) = K_{4/3}(x)K_{1/3}(x)-\frac{3}{5}x[K_{4/3}^2 (x)-K_{4/3}^2(x)]
\end{equation*}

Here, $K_n(x)$ are the modified Bessel functions, $x_R = \frac{h\nu}{h\nu_R}$ $\nu_R = \frac{3e}{4 \pi m_e c}\gamma^2 B_R$, and $\nu$ is the frequency of the emitted photon. 

For an ordered magnetic field $B_O$, the pitch angle integrated spectrum can be written as 

\begin{equation}
  N_O(x_O) = \frac{2 \pi e^2 }{\sqrt{3} \gamma^2 c h} \int_{x_O}^{\infty} K_{5/3} (x_{O}') dx_O'
  \label{spec_order}
\end{equation}
where, $x_O = \frac{h\nu}{h\nu_O}$ and $\nu_O = \frac{3 e}{4 \pi m_e c} \gamma^2 B_O sin\theta$. Here $\theta$ is the angle between the line of sight and the magnetic field. The above expression is the same as the standard expression for a line of sight integrated spectrum from a particle moving with a pitch angle $\alpha$, with $\theta$ replaced by $\alpha$ \citep[e.g.][]{1986rpa..book.....R}. Since, in its derivation it is assumed that the contribution to the spectrum arises only when the angle between the line of sight and pitch angle is small ($< 1/\gamma$), the resultant expression is the same, when one integrates over the line of sight ($\theta$) or over the pitch angle ($\alpha$). The total power for a random distribution $P_R$ can be obtained by integrating Equation \ref{spec_rand} over frequency. The total power for an ordered field $P_O$, can be obtained by averaging Equation \ref{spec_order} over $\theta$ and integrating over frequency. We have verified that $P_R = (2/3) P_O$ as expected \citep{1986rpa..book.....R} when $B_R = B_O$.

To obtain the observed spectrum from a jet at a redshift, $z$ and having a Doppler factor, $\delta_D$, the single particle spectra need to be Doppler boosted and integrated over a particle distribution $n_\gamma (\gamma) $. As shown by \cite{Hota_2021}, it is convenient to transform $\gamma$ to another variable,  $\xi=\mathbb{\sqrt{C} \gamma}$, where 
$\mathbb{C}=(\frac{3eh}{4 \pi m_e c})\frac{\delta_D B_R}{(1+z)}$. Note that with transformation, $\xi^2$ has units of energy and corresponds approximately to the photon energy where the single particle spectrum peaks. With this transformation, the predicted synchrotron photon spectrum, apart from an overall normalization, depends only on the form of the particle energy distribution, $n(\xi)$ and not on degenerate parameters such as $\delta_D$ and $B_R$.

For a system, which has an ordered and random field, the observed spectrum will be the sum of the two components,

\begin{equation}
    S_T(E) = S_R(E) + S_O(E)
\end{equation}
\label{spec1}
where,

\begin{equation*}
    S_R(E) = \frac{\delta_D^3(1+z)V}{d_L^2\sqrt{\mathbb{C}}} \int_{\xi_{min}}^{\xi_{max}} (1-f_O) N_R (\frac{E}{\xi^2}) n(\xi) d\xi
\end{equation*}
and 
\begin{equation*}
    S_O(E) = \frac{\delta_D^3(1+z)V}{d_L^2\sqrt{\mathbb{C}}} \int_{\xi_{min}}^{\xi_{max}} f_O N_O (\frac{E}{\eta \xi^2}) n(\xi) d\xi
\end{equation*}

where $E$ is the photon energy,  $\eta = \frac{B_O sin\theta}{B_R}$ and $f_O$ is the fraction of the volume, V, that can be associated with the ordered field. 

\noindent The polarization degree $\Pi_O$ from the ordered field is \citep{1986rpa..book.....R} 

\begin{equation*}
    \Pi_O(E) = \frac{\int_{\xi_{min}}^{\xi_{max}} K_{2/3} (\frac{E}{\eta \xi^2})f_o n(\xi) d\xi}{\int_{\xi_{min}}^{\xi_{max}} \left[ \int_{E/\eta \xi^2}^{\infty} K_{5/3} (x') dx' \right] f_o n(\xi) d\xi}
\end{equation*}
Thus the observed polarization degree is, 

\begin{equation}
    \Pi_T(E) = \frac{S_O(E)}{S_T{(E)}} \Pi_O(E)
    \label{pol}
\end{equation}

\begin{table*}
\normalsize
\centering
\renewcommand{\arraystretch}{1.5} 
\caption{\label{tab:data} Summary of \textit{NuSTAR}, \textit{Swift}-XRT/UVOT and \textit{IXPE} observations of Mrk 501 for MJD = 59665.}
\setlength{\tabcolsep}{2pt}
\begin{tabular}{l c c c c r}
\hline
 & ~~~ \textit{NuSTAR}~~~ & ~~~ \textit{Swift}-XRT/UVOT ~~~ & ~~~ \textit{IXPE} \\
\hline
Obs ID & 60702062004 & 00011184187 & 01004601 \\
Date and Time ~~~ & ~~~ 2022-03-27 T01:26:09 ~~~ & ~~~ 2022-03-27 T03:49:36 ~~~ & ~~~ 2022-03-27 T05:39:23 \\
Exposure (ks) & 20.29 & 0.93/0.92 & 86 \\
\hline
\end{tabular}
\end{table*}

\section{Polarization of Mrk 501} \label{sec:sec3}

On 27th March 2022, the degree of X-ray polarization ($11.5\pm 1.6$\%) was significantly higher than the median intrinsic optical polarization degree, $5\pm 1\%$\citep{2022Natur.611..677L}. The optical polarization degree estimate is the median value obtained from several measurements from different telescopes and has been corrected for host galaxy contamination. There were near-simultaneous observations undertaken by {\it NuSTAR} and {\it Swift}-XRT and UVOT. The details of these are listed in Table \ref{tab:data}. The details of the spectral extraction and analysis of these data are given in \cite{10.1093/mnras/stae706} and hence we do not repeat them here. We do not consider the higher energy \textit{Fermi}-LAT data since the focus here is on the synchrotron emission. For different particle energy distributions, we formally fit the optical and X-ray spectra, and the polarization values using the formulation described in the earlier section, with emphasis on the polarization degree predictions. The polarization degree as a function of energy is converted to an Xspec \citep{1996ASPC..101...17A} readable format and then both the spectral energy distribution and polarization degree are fitted within the Xspec environment. A local Xspec convolution model was created such that if a flag is set to one, then the output is the broadband synchrotron spectrum (using Equation \ref{spec1}), and if set to zero it outputs the polarization degree (using Equation \ref{pol}). The model is then convolved with different particle distributions to obtain a fit with parameter values and errors. As mentioned below, the spectral energy distribution is rather insensitive to the parameters representing the ordered magnetic field, namely $f_O$ and $\eta$. Hence, the spectral fitting is the same as those reported in \cite{10.1093/mnras/stae706}.

\subsection{Particle Distributions}

\begin{figure*}
\centering
    \includegraphics[width=.5\textwidth]{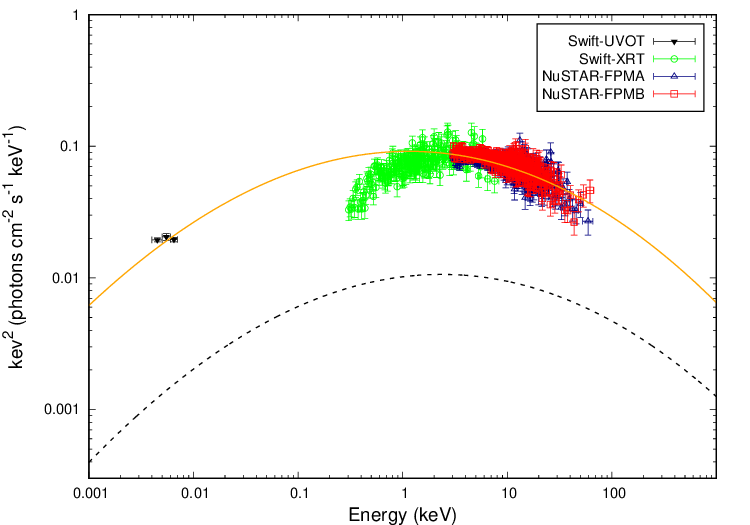}\hfill
    \includegraphics[width=.5\textwidth]{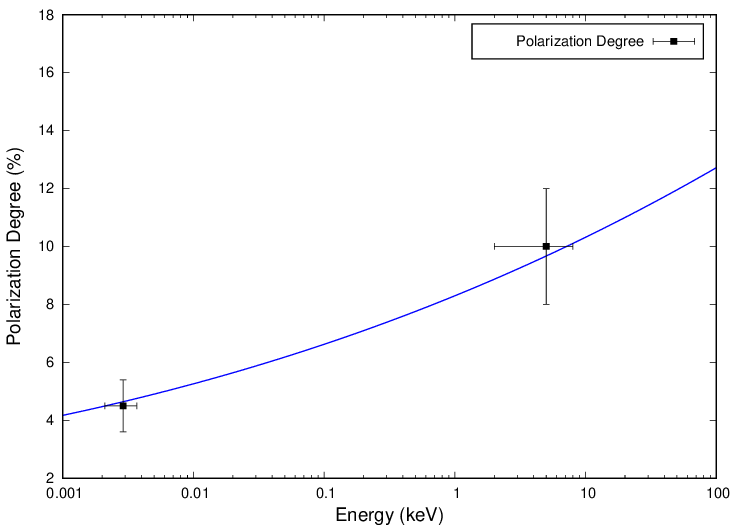}
    \includegraphics[width=.5\textwidth]{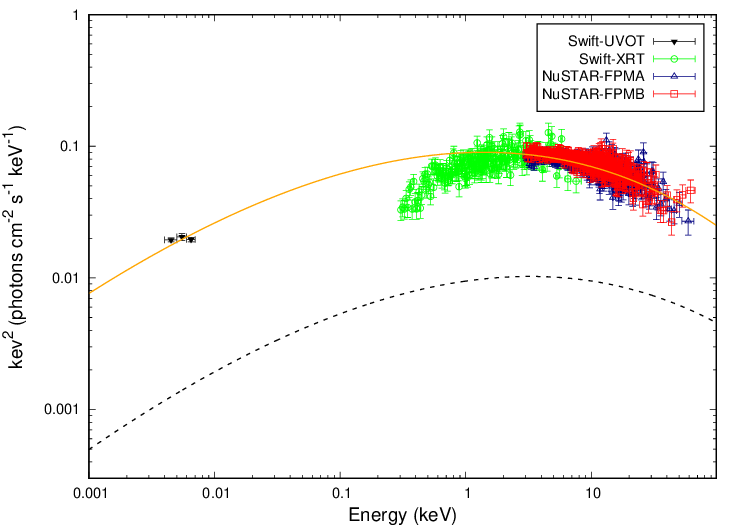}\hfill
    \includegraphics[width=.5\textwidth]{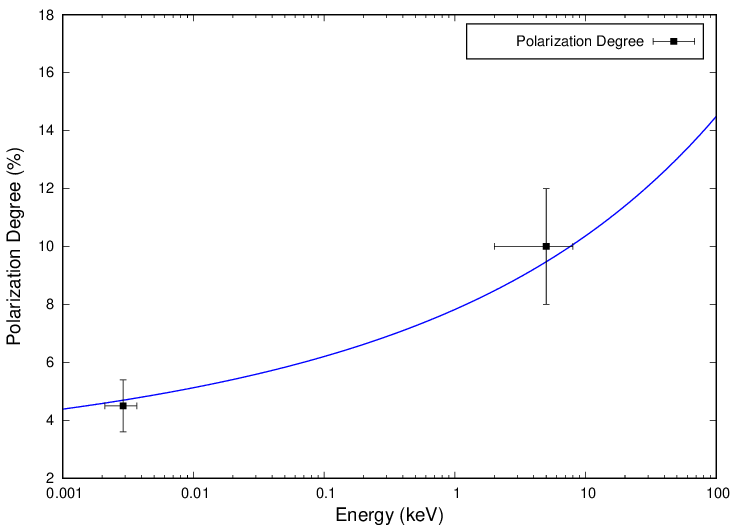}
    \caption{The data points in the left panels show the broadband SED, while the right panels show the polarization degree in the X-ray and optical bands \citep{2022Natur.611..677L} for Mrk 501. Note that the model represents an unabsorbed spectrum while the X-ray photon spectra are the absorbed one for log parabola (top panel) and EDA (bottom panel) particle distributions. The UV data points have been dereddened using E(B-V) = 0.017 mag \citep{10.1093/mnras/stae706}. The dotted lines are the synchrotron emissions from the ordered magnetic field components. The solid lines in the right panels show the corresponding expected variation of polarization degree.}
    \label{fig1}
    
\end{figure*}

We first consider the energy distribution of the particles to be a log parabola such that,
\begin{equation}\label{lp}
    n({\xi})=K \left (\frac{\xi}{\xi_r} \right)^{- \alpha - \beta \text{log} \left(\frac{\xi}{\xi_\text{ref}} \right)}
\end{equation}

where $\alpha$ is the index at the reference energy $\xi_\text{ref}$ which is fixed at $1 \sqrt{\text{keV}}$, $\beta$ is the curvature parameter and $K$ is the normalization factor. The spectral fitting is shown in the top left panel of Figure \ref{fig1}, with the best-fit parameters, $\alpha = 2.9 $ and  $\beta = 0.59$. The solid line represents the total best-fit spectrum $S_T(E)$. To fit the X-ray data the Galactic absorption has been taken into account using the XSPEC model {\it Tbabs}  with column density fixed at $ 1.69 \times 10^{20}$ cm$^{-2}$ \footnote{\url{https://heasarc.gsfc.nasa.gov/cgi-bin/Tools/w3nh/w3nh.pl}}. Note that the model shown in the Figure \ref{fig1} is the unabsorbed one and hence there is a deviation from the data at low energies. The UV data presented is the derredened one \citep{10.1093/mnras/stae706}. The dotted line represent the emission from the ordered field $S_O (E)$, for an ordered field volume fraction $f_O = 0.022 $ and ratio $\eta = \frac{B_O sin\theta}{B_R} = 1.95 $. The top right panel of the Figure \ref{fig1}, shows the polarization degrees in the X-ray and optical as reported by \cite{2022Natur.611..677L}, with the solid line being the predicted energy polarization degree $\Pi_T(E)$. Since the detected polarization degree is small, the spectral component from the ordered magnetic field $S_O(E)$ is required to be significantly smaller than the one from the random field $S_R(E)$ and hence the total spectrum $S_T (E)$ is rather insensitive to $f_O$ and $\eta$. The polarization degree data points constrain the values of $f_O$ and $\eta$.

We also consider another particle distribution which arises if the acceleration time scale is considered to have an energy dependence. This distribution which is described in detail in \cite{Hota_2021}, \cite{Khatoon_2022}, \cite{10.1093/mnras/stae706}, \cite{hota2024multiwavelengthstudyextremehighenergy} and \cite{tantry2024probingbroadbandspectralenergy}. This distribution is called Energy Dependent Acceleration (EDA) model and is given as,
\begin{equation}\label{eda}
         n(\xi)=K(\xi)^{k-1} \text{exp} \left [-\frac{\psi}{k}\xi^{k} \right] 
\end{equation}
where $\psi$ and $\kappa$ are parameters. The bottom panels of Figure \ref{fig1} show the spectra and polarization degree while the best-fit parameters are listed in Table 2.  The parameters $\xi_0$ and $\kappa$ are related to the dependence of the acceleration time-scale on energy and their physical interpretation has been discussed in earlier works \citep{2020MNRAS.499.2094G, Hota_2021, Khatoon_2022, 10.1093/mnras/stae706}. We have included such a particle distribution in this work,  to highlight that intrinsically curved particle energy distributions (which are not necessarily of the log-parabola form) can reproduce the observed energy-dependent polarization.

\begin{figure*}
\centering
    \includegraphics[width=.5\textwidth]{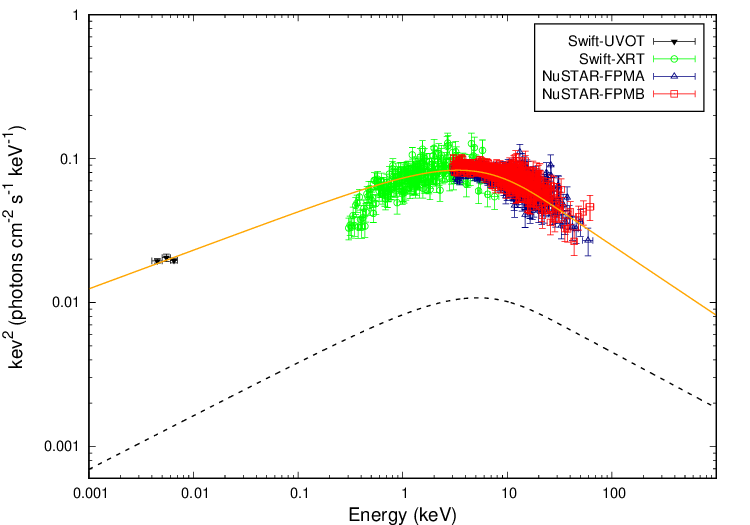}\hfill
    \includegraphics[width=.5\textwidth]{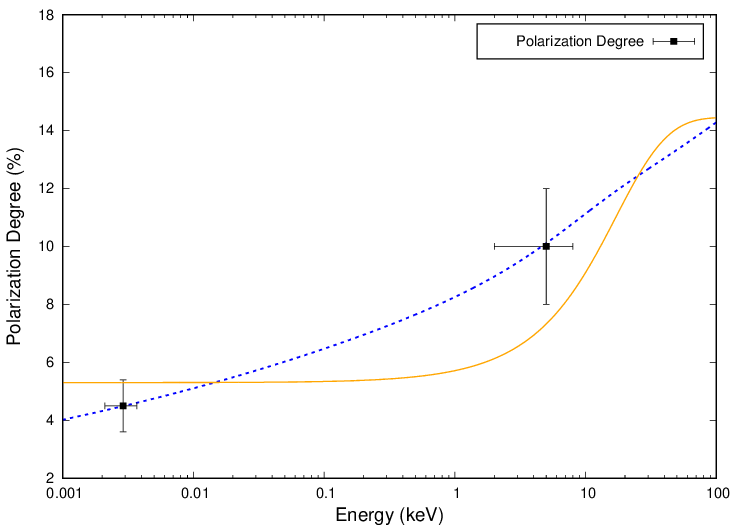}\hfill
    \caption{Broadband SED (left panel) and polarization degree data points (right panel) for Mrk 501. The solid line in the right panel shows the closest expected polarization degree which does not match with the data. The dotted line in the right panel is the expected polarization degree when the ordered field volume factor is assumed to increase with particle energy and the left panel shows the corresponding best fit total unabsorbed total photon spectrum (solid line) and the ordered field component (dotted line).  }
    \label{fig2}
\end{figure*}

We next consider the case when the particle energy distribution is taken to be a broken power law of the form,

\begin{equation}\label{bpl}
    n(\xi)=
\begin{cases}
    K (\xi/1\sqrt{{\text{keV}}})^{-p_1} & \text{for} \hspace{3pt} \xi < \xi_\text{break} \\
    K \xi^{p_2-p_1}_\text{break}(\xi/1\sqrt{{\text{keV}}})^{-p_2} & \text{for} \hspace{3pt} \xi > \xi_\text{break}  
\end{cases} 
\end{equation}
where $p_1$ and $p_2$ are the indices before and after the break energy $\xi_{brk}$. The right panel of Figure \ref{fig2} shows that the spectrum with $p_1 = 2.46 $, $p_2 = 3.99 $ and $\xi_{brk}^2 = 4.97 $ keV. The solid line in the left panel shows the expected polarization for $f_o = 0.0075$ and $\eta = 3.17$, which are the values which bring the expected polarization closest to the observed ones. As reported by \cite{2022Natur.611..677L}, the higher polarization measured in the X-ray compared to the optical cannot be explained. Due to the change in the particle index, the expected polarization degree increases for high energies beyond the break energy, but cannot explain the observed low energy X-ray polarization. As proposed by \cite{2022Natur.611..677L} the solution maybe that the high-energy electrons experience a more ordered field than the low energy ones. In the formulation presented here, this is equivalent to ordered volume fraction $f_o$ to be energy dependent. For simplicity, we could choose this dependence to be a power law such that $f_o(\xi) = f_{ox} \xi^\chi$. Using this form the observed polarization degree maybe explained as shown in the dotted line in the right panel of Figure \ref{fig2} with the best-fit parameters listed in Table \ref{tab:bestfit}. Since there are now three parameters to explain two polarization degree points, we can fix $\eta = 1$ and find $f_{xo} = 0.076 $ and $\chi = 0.21$. As mentioned earlier the spectral fitting is rather insensitive to these values. Thus a modest dependence of $f_o$ on energy can also explain the observations.

\begin{table*}
\caption {\label{tab:bestfit} The best-fit parameters obtained for all the particle distributions considered. For BPL, the $\eta = \frac{B_O sin \theta}{B_R}$ is frozen at 1. The errors have been calculated at 90\% confidence level.}
\renewcommand{\arraystretch}{1.55} 
\normalsize
\centering
\begin{tabular}{l c c c c c r}
\hline
\multicolumn{7}{c}{\textbf{Log Parabola}} \\  
\hline
 & $\alpha$ \quad \quad \quad \quad & \quad \quad \quad \quad $\beta$ \quad \quad \quad\quad & \quad \quad \quad \quad $\eta$ \quad \quad\quad \quad &  \quad \quad \quad \quad $f_{o}$  \quad \quad\quad \quad & \quad \quad\quad \quad $\chi^{2}_\text{red}(\text {dof})$ \\
 & 2.92$^{+0.02}_{-0.02}$ & \quad \quad 0.59$^{+0.02}_{-0.02}$ & \quad \quad 1.95$^{+1.4}_{-0.9}$ &  \quad \quad 0.022$^{+0.02}_{-0.01}$ & \quad \quad \quad \quad 1.17(948)\\ 
\hline
\multicolumn{7}{c}{\textbf{Broken Power Law}} \\  
\hline
 & $p_1$ & $p_2$ & $\xi_\text{brk}$ $\sqrt{\text{keV}}$ & $\chi$ & $f_{x o}$ & $\chi^{2}_\text{red}(\text {dof})$ \\
 & 2.46$^{+0.02}_{-0.02}$  & 3.99$^{+0.10}_{-0.13}$ &  2.23$^{+0.15}_{-0.14}$ & 0.21$^{+0.15}_{-0.15}$ & 0.076$^{+0.02}_{-0.02}$ & 0.98 (947)\\ 
\hline
\multicolumn{7}{c}{\textbf{EDA}} \\  
\hline
& $\psi$ & $k$ & $\eta$ & $f_{o}$ & $\chi^{2}_\text{red}(\text {dof})$ \\
& 2.10$^{+0.03}_{-0.02}$ & 0.25$^{+0.01}_{-0.01}$ &  2.24$^{+2.15}_{-1.17}$ & 0.01$^{+0.04}_{-0.005}$ & 1.02 (947)\\ 
\hline
\end{tabular}

\end{table*}

\section{Discussion \& Conclusions}\label{sec:Discussion}

It is shown in this work, that if the particle energy distribution has intrinsic curvature, the broadband spectral energy distribution of Mrk 501, as well as the polarization degree measured in X-ray and optical bands can be explained for a system where all the particles experience the same magnetic configuration. On the other hand, as reported by \cite{2022Natur.611..677L}, if the particle distribution is a power law (or a broken power law) both the spectrum and polarization can be reproduced if the higher energy particles exist in a region of more ordered magnetic field compared to the low energy ones. The question then is how one can distinguish between these two scenarios?

Theoretical studies may be undertaken to consider whether particles with energy distributions that are intrinsically curved can be generated in physically reasonable situations \citep[e.g.][]{Kundu_2021, 2023ApJ...952....1D}. Similarly, particle diffusion may need to be studied or simulated to understand whether higher energy particles would preferably exist in regions of more ordered magnetic field \citep[e.g.][]{2008A&A...488..795R, 2017ApJ...845...43K}. However, given the complexities and uncertainties of the acceleration and diffusion processes, a clear discrimination between the scenarios may not be forthcoming.

High quality broadband spectral data may be able to distinguish whether the underlying particle distribution has curvature or not. However, the available spectral data is not sufficient to make such distinctions, especially since the spectral models are often degenerate \citep{Hota_2021, 10.1093/mnras/stae706, hota2024multiwavelengthstudyextremehighenergy, tantry2024probingbroadbandspectralenergy}. Other considerations such as correlations between spectral parameters \citep{Hota_2021} and minimum jet power requirements \citep{10.1093/mnras/stae706} may provide some clues, but are not strictly conclusive.

It is not expected in the near future, to have significantly better X-ray polarization measurements and to have measurements in different energy bands. The presently available photon energy-dependent polarization data (even with some factor of few improvements) may not be able to distinguish between different models, given that one can adjust parameters to fit the data. Perhaps a more promising approach would be to study the temporal variability of the polarization and correlate them with spectral parameters. The three \textit{IXPE} observations of the source do not show any significant variability in the polarization degree for the first two observations, while for the last one there is marginal decrease in both X-ray and optical and moreover this observation does not have simultaneous \textit{NuSTAR} data \citep{2024A&A...685A.117M}. Attempts have been made to study correlations between X-ray spectral parameters and polarization using four IXPE observations of Mrk 421 \citep{2023NatAs...7.1245D, 2024A&A...681A..12K}. Given the uncertainty in the X-ray polarization measurement for a single observation, a number of observations are required with simultaneous spectral coverage during different spectral states. For example, the ratio between the X-ray and optical polarization degree would be expected to increase with increasing particle distribution curvature (i.e. $\beta$) and any observed correlation between the two would support the interpretation that the particle distribution indeed has curvature. Unfortunately, the absence of such a correlation may not rule out the model since the polarization degree ratio can also vary due to variations in $\eta$ i.e. ratio between the magnitude of the ordered and random magnetic field. Nevertheless, since polarization degree is closely linked to the underlying shape of the particle distribution, a number of polarization measurements along with simultaneous spectral data would be important to understand the nature of these systems. 

\section*{Acknowledgements}
We acknowledge the use of public data from the \textit{NuSTAR}, \textit{Swift}-XRT/UVOT from NASA’s High Energy Astrophysics Science Archive Research Center (HEASARC). R.M. thanks Prof. Shiv Sethi for useful discussions on the expected polarization from Synchrotron emission. H.B. and R.G. would like to acknowledge IUCAA for their support and hospitality through their associateship program.

\bibliography{sample631}{}
\bibliographystyle{aasjournal}



\end{document}